\documentclass[12pt]{iopart}
\usepackage{graphicx}
\usepackage{dcolumn}
\usepackage{bm}% bold math

\usepackage{colordvi}

\begin{document}

\title[Engineering ultralong spin coherence in two-dimensional hole systems]{Engineering ultralong spin coherence in two-dimensional hole systems at low temperatures}
\author{T Korn, M Kugler, M Griesbeck, R Schulz, A Wagner, M Hirmer, C Gerl, D Schuh, W Wegscheider\footnote{present address: Solid State Physics Laboratory, ETH Zurich, 8093 Zurich, Switzerland} and C Sch\"uller}
\ead{tobias.korn@physik.uni-regensburg.de}
\address{Institut
f\"ur Experimentelle und Angewandte Physik, Universit\"at
Regensburg, D-93040 Regensburg, Germany}

\date{\today}

\begin{abstract}
For the realisation of scalable solid-state quantum-bit systems, spins in semiconductor quantum dots are promising candidates.
A key requirement for quantum logic operations is a sufficiently
long coherence time of the spin system. Recently, hole spins in
III-V-based quantum dots were discussed as alternatives to electron spins, since the hole spin, in contrast to the electron spin, is not affected by contact hyperfine interaction with the nuclear spins.
Here, we report a breakthrough in the spin coherence times of hole ensembles, confined in so called natural quantum dots, in narrow GaAs/AlGaAs quantum wells at temperatures below 500~mK. Consistently, time-resolved Faraday rotation and resonant spin amplification techniques deliver hole-spin coherence times, which approach in the low magnetic field limit values above 70~ns. The optical initialisation of the hole spin polarisation, as well as the interconnected electron and hole spin dynamics in our samples are well reproduced using a rate equation model.
\end{abstract}
\submitto{\NJP}
\maketitle

\section{Introduction}
Among the most promising systems for the realisation of quantum
computing devices are spins in semiconductor quantum dots (QDs)
\cite{Awschalom_QC}. Using electrostatic gating techniques, these
dots may be defined within a two-dimensional electron system (2DES)
by local depletion. This approach has the advantage that it allows
for the fabrication of scalable arrays of quantum bits, as it is
based on high-resolution lithography techniques instead of
self-organized growth of QDs. While the spin dephasing of electrons
in high-mobility GaAs/AlGaAs-based 2DES is extremely fast - on the
order of only a few tens of picoseconds \cite{Brand,Stich1,Stich2}
in the low-excitation limit - for electrons confined in QDs, spin
dephasing is strongly reduced and the main remaining spin dephasing
channel is contact hyperfine interaction with the nuclei
\cite{Loss02}. The latter may be suppressed by using elaborate spin
echo techniques \cite{Petta05}. Several alternative material systems
like silicon \cite{Eriksson04} and graphene \cite{Trauzettel07} have
been suggested to overcome the problem of hyperfine interaction of
electrons and nuclei. Recently, hole spins confined in QDs have been
considered for quantum computing, as long hole spin dephasing times
(SDT) were observed in p-doped self-organized QDs
\cite{heiss,Marie09}. Additionally, efficient electrical tuning of
the effective hole $g$ factor in low-dimensional structures was
predicted \cite{andlauer09} and experimentally demonstrated
\cite{Kugler09}. In the present work we demonstrate that very long
hole SDTs can be observed in a two-dimensional hole system (2DHS),
residing in a narrow quantum well (QW) in a GaAs/AlGaAs-based
heterostructure. We show that the hole SDT strongly depends on the
energy splitting between the quantized heavy-hole (HH) and
light-hole (LH) states within the QW, which is controlled by the QW
width. Specifically, the hole SDT can reach values above 70~ns at
low temperatures and in small magnetic fields, which is about two
orders of magnitude longer than previously reported results on wider
GaAs QWs \cite{Kugler09,syperek}. In particular, these long SDTs
allow us to use resonant spin amplification (RSA) techniques for
precise measurements of the hole spin dynamics. The results shown
here suggest that GaAs-based 2DHS may be a viable alternative to
2DES for quantum computing applications, which rely on
electrostatically confined charged carriers.

\section{Sample design and experimental methods}
Our samples are single-side p-modulation-doped
GaAs/Al$_{0.3}$Ga$_{0.7}$As QWs containing a 2DHS with relatively
low hole density (see Table \ref{Data}). Previous investigations on
similar samples with relatively wide wells (15~nm and 10~nm)
\cite{Kugler09} showed that at these low densities, the optical
recombination spectra at liquid-Helium temperatures are governed by
recombinations of neutral and positively-charged excitons, i.e.,
bound excitonic complexes, consisting of one electron and two holes.
Most importantly, even for the wider QWs, at very low temperatures,
the resident holes become localised in potential fluctuations in the
plane of the QW \cite{Kugler09}. It was shown by Syperek et
al.~\cite{syperek} and Kugler et al.~\cite{Kugler09} that
localisation of the holes is crucial for the observation of long
hole SDTs on the order of a few-hundred picoseconds in those
samples. Here, we report on hole SDTs in narrow QWs, which are up to
two orders of magnitude longer.

The structures are grown by molecular-beam epitaxy (MBE) on [001]
substrates. Some characteristic properties are listed in Table
\ref{Data}. The table clearly shows how the hole mobility is
significantly reduced for the  thinnest sample D, most
likely due to the increased influence of monolayer fluctuations at
the AlGaAs/GaAs interfaces on the hole wave function within the QW.
\begin{table}
\caption{{\bf Sample data.} Density and mobility were determined
from
 magnetotransport measurements at 1.3~K.}
\begin{indented}
\item[]
\lineup
\begin{tabular}{@{}lllll}
  \br
  Sample& QW width & hole density $p$  & hole mobility $\mu$ & electron g \\
    & (nm) & $(10^{11}$~cm$^{-2})$ & ($10^{5}$~cm$^2/$Vs) &  factor $|g_e|$\\
  \mr
  A &  15 & 0.90 & 5.0 & $0.280 \pm 0.005 $\\   % D040817B
  B &  9 & 1.03 & 3.6 & $ 0.133 \pm 0.01 $\\         % D080808A-1 (9 nm)
  C & 7.5 & 1.10 & 5.3& $ 0.106 \pm 0.01 $\\           % D080430A-8 (7,5 nm)
  D &  4 & 1.10 & 0.13  &$ 0.266 \pm 0.003 $\\ % D090210C
  \br
\end{tabular}
\end{indented}
 \label{Data}
\end{table}
For time-resolved Faraday rotation (TRFR) measurements, the samples
are first glued onto a sapphire substrate with optically transparent
glue, then the semiconductor substrate is removed by grinding and
selective wet etching. All samples contain a short-period GaAs/AlGaAs
superlattice, which serves as an etch stop, leaving only the
MBE-grown layers. The TRFR and resonant spin amplification (RSA) measurements are performed in an
optical cryostat with $^3$He insert, allowing for sample
temperatures below 400~mK and magnetic fields of up to 11.5~Tesla. A
pulsed Ti-Sapphire laser system generating pulses with a length of
600~fs and a spectral width of 3-4~meV is used for the optical
measurements. The repetition rate of the laser system is 82~MHz. The
laser pulses are split into a circularly-polarised pump beam and a
linearly-polarised probe beam by a beam splitter. A mechanical delay
line is used to create a variable time delay between pump and probe.
Both beams are focussed to a diameter of about 80~$\mu$m on the
sample using an achromat, resulting in an excitation density of about 2~Wcm$^{-2}$. The center wavelength of the laser system
is tuned to near-resonance with the HH excitonic absorption lines of
the samples. For this, photoluminescence spectra, taken using
nonresonant excitation, are used to determine the transition
energies of the neutral and charged excitons~\cite{Kugler09}. As the spectral linewidth of our laser system exceeds the energy splitting between the charged and
neutral exciton transitions, both are excited by the pump laser pulses. In the
TRFR and RSA experiments, the circularly-polarised pump beam is
 generating electron-hole pairs in the QW, with spins
aligned parallel or antiparallel to the beam direction, i.e., the QW
normal. Due to the spectral linewidth of the laser, typically, both,
neutral and positively charged excitons are excited near-resonantly.
In the TRFR measurements, the spin polarisation created
perpendicular to the sample plane by the pump beam, is probed by the
time-delayed probe beam via the Faraday effect: the axis of linear
polarisation of the probe beam is tilted by a small angle~\cite{angle}, which is
proportional to the out-of-plane component of the spin polarisation.
This small angle is detected using an optical bridge. A lock-in
scheme is used to increase sensitivity. In the RSA measurements, the
Faraday rotation angle is measured for a fixed time delay as a
function of an applied in-plane magnetic field. In our
investigations, we exploit the strengths of both methods in order to
explore the limits of hole spin dynamics in GaAs-based hole systems:
In the TRFR experiments, one measures in the time domain via a
pump-probe scheme, and the SDT as well as photocarrier lifetimes can
be extracted from the measurements. The disadvantage of this method,
however, is the limitation of the accessible time range by the
travel length of the optical delay line. In our setup, this limits
the time range to about 2~ns. This hinders an accurate determination
of SDTs, which are significantly longer than the measurement range.
The TRFR method encounters severe problems, if the SDT is even
longer than the time interval between two subsequent laser pulses in
the pulse train of the mode-locked laser (in our case about 12~ns).
Fortunately, with the RSA technique \cite{Kikkawa98}, one can
overcome these problems, since here one works with a fixed time delay
between pump and probe pulses. The method is based on the resonant
amplification of the Faraday signal, when integer multiples of the
precession period of the spins in an inplane magnetic field coincide
with the inverse laser repetition rate, i.e., the time interval
between subsequent pulses. However, here the extraction of the SDT
from the experimental data is more involved, since it is hidden in
the linewidths of the RSA maxima. As we will show below, in
particular in our case, where electron and hole recombination and
spin dynamics are interconnected, this is rather complex.

\section{Quantum-well width dependence of hole spin-dephasing time}
We start our discussion by presenting TRFR experiments on our
samples as a function of an in-plane magnetic field. As discussed
above, in these experiments we are detecting the charge carrier and
spin dynamics of photoexcited electrons and  holes within a time
range $<2$~ns. This will allow us to gain important information on
the interconnected electron and hole photocarrier and spin dynamics.
\begin{figure}
\centering
  \includegraphics[width= 0.8\textwidth]{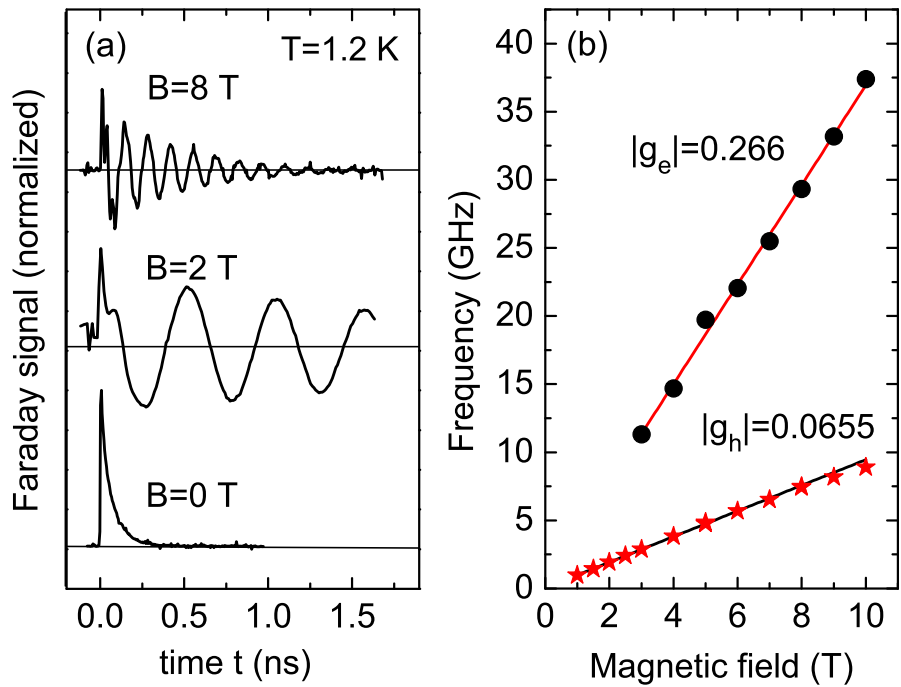}
   \caption{{\bf TRFR for different in-plane magnetic fields and dispersion of precession frequencies.} (a) TRFR traces for sample D, measured at 1.2~K for different magnetic fields.
   (b) Dispersion of electron (black dots) and hole (red stars) spin precession determined for sample D from TRFR measurements. The solid lines are linear fits to the data.}
   \label{4nm_2Panel}
\end{figure}
Figure \ref{4nm_2Panel}(a) shows a series of TRFR traces measured on
sample D, the sample with the thinnest QW, at 1.2~K at three
different magnetic fields, applied in the sample plane. The trace at
$B=0$ shows a single exponential decay of the Faraday signal after
pulsed excitation with a decay time of $\tau_R$= 70~ps. Each of the
other two traces, however, exhibits the superposition of two damped
oscillations with markedly different frequencies and damping
constants: A fast oscillation, which is observable within about the
decay time of the $B=0$ trace, only, and a slower oscillation,
which, in the $B= 2$~T trace, exceeds the measurement range by far.
The sum of two damped cosine functions is fit to the data in order
to extract the precession frequencies and decay constants. As a
result, we can identify the high-frequency oscillations, which decay
significantly faster, as the precession of photogenerated electrons,
and the low-frequency oscillations as the precession of resident
holes within the sample: In figure~\ref{4nm_2Panel}(b), the dispersions
of the two precession frequencies are plotted, and the electron and
hole $g$ factors are extracted from the data. The electron $g$
factor $|g_e| =0.266$ is in good agreement with values measured for
QWs of similar widths~\cite{Snelling91,Yugova_g}, while the hole $g$ factor
$|g_h|=0.0655$ is close to zero, as was theoretically predicted for
GaAs-based QWs grown along the [001] direction \cite{winkler} and
experimentally observed \cite{Kugler09,syperek,marie99}.

We note that the decay constant of the electron spin precession,
$\tau_e = 75 \pm 5$~ps, remains almost constant throughout the
investigated magnetic field range. This may be explained as follows: as
our samples are p-doped, the optically oriented electron spins can
only be observed during the photocarrier lifetime. The measured
decay constant $\tau_e = 75 \pm 5$~ps therefore corresponds to the
photocarrier lifetime in sample D, $\tau_R$, as the electron SDT is typically
longer. In stark contrast, the hole SDT decreases strongly with
increasing magnetic field, as can be seen quite drastically from the
$B=2$~T and $B=8$~T traces in figure~\ref{4nm_2Panel}(a). However, at
this stage the important question arises, why the trace at $B=0$ in
figure~\ref{4nm_2Panel}(a) obviously does not show a long lasting hole
spin signal but a spin signal, which is governed by the photocarrier
lifetime, only? We will address this question further below in the
next section.

\begin{figure}
\centering
  \includegraphics[width= 0.8\textwidth]{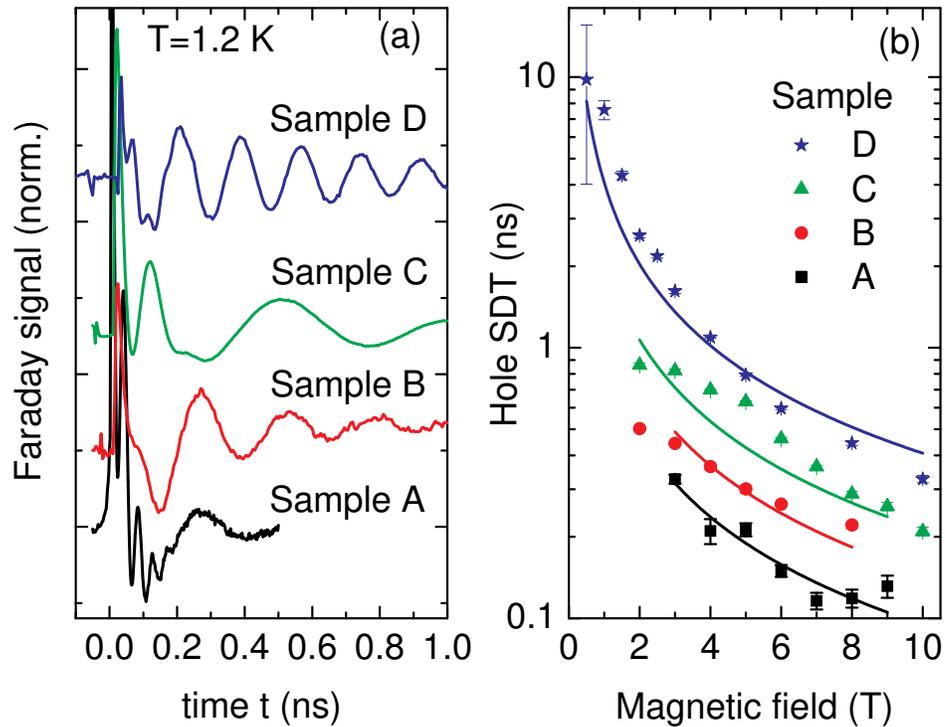}
   \caption{{\bf TRFR for different well widths and magnetic-field dependence of SDTs.} (a) TRFR traces for samples A, B, C and D at 1.2~K.
   A 6~Tesla   in-plane magnetic field was applied during the measurements.
   (b) Hole spin dephasing times for samples A, B, C and D as a function of magnetic field.
   The lines represent fits of a 1/B dependence.}
   \label{Vgl_Width_2Panel}
\end{figure}

To investigate the strong magnetic field dependence of the hole SDT
in more detail, we study the hole spin dynamics as a function of the
applied in-plane magnetic field for samples with different QW
widths. Figure~\ref{Vgl_Width_2Panel}(a) shows TRFR traces for all 4
samples, measured at 1.2~K with an applied in-plane magnetic field
of 6~T. The traces have been scaled and vertically shifted in order
to allow easy comparison of the hole spin dynamics. In all traces,
again, a fast, rapidly decaying electron spin precession can be
observed within the first $\sim 100$~ps. After photocarrier
recombination, only larger-period hole spin precession is visible in
the traces. Even though the applied in-plane magnetic field was
fixed at 6~T, in all measurements shown in
figure~\ref{Vgl_Width_2Panel}(a), it can be seen clearly that both, the
electron and the hole spin precession frequencies, are different for
all 4 samples. For electrons, it is
well-known that the  $g$ factor in a QW depends on the QW
width~\cite{Snelling91,Yugova_g}: due to the nonparabolicity of the conduction band in GaAs, the electron g factor changes depending on the electron energy above the conduction band edge, and in good approximation, the QW confinement energy of an electron will lead to a similar change of the g factor. Additionally, as the electron wave function has a nonvanishing amplitude within the QW barriers for thin GaAs QWs, the different electron $g$ factors of pure GaAs and the barrier material are  admixed. The values of the electron $g$ factors extracted from TRFR measurements on all four samples are
listed in Table \ref{Data}. We note that the sign of the g factor cannot be directly determined from TRFR measurements, however, by comparing our measured values to literature data~\cite{Snelling91,Yugova_g}, we conclude that the electron g factor for samples A-C is negative, while it is positive for sample D.

On the other hand, the different hole $g$ factors observed in the
TRFR traces stem from the strong anisotropy of the hole $g$ tensor,
which was predicted by Winkler et al. \cite{winkler} for [001]-grown
GaAs QWs. The effective hole $g$ factor, $g_h^{*}$, measured in TRFR
is given by the geometric sum of the in-plane (${g}_{\bot}$) and
out-of-plane (${g}_{||}$) components of the hole $g$ tensor:
\begin{equation} {g}_{h}^{*} =
\sqrt{{g}_{\bot}^{2}cos^{2}\alpha
    + {g}_{||}^{2}sin^{2}\alpha}.
\label{GeoSum}
\end{equation}
Here, $\alpha$ is the tilt angle of the magnetic field with respect
to the QW plane. While the in-plane component of the hole $g$ factor
is close to zero, ${g}_{\bot}\sim 0$, the out-of-plane component is
${g}_{||} \sim -0.7$ for bulk GaAs~\cite{landolt-boernstein}. Therefore, even small tilt
angles $\alpha$ result in markedly different $g_h^{*}$.

In figure~\ref{Vgl_Width_2Panel}(b), the hole SDTs for all 4 samples are
shown as a function of the applied magnetic field. Two effects are
clearly visible here:
\begin{enumerate}
  \item[(i)] For all samples, the hole SDT decreases as the magnetic field is increased, following approximately a $1/B$ dependence.
  \item[(ii)] The hole SDT increases systematically as the QW width is reduced. It is smallest for sample A with the widest QW and largest for sample D with the thinnest QW.
\end{enumerate}

The decrease of the hole SDT with magnetic field is a well-known
phenomenon. It is believed to be caused by the inhomogeneity of the
hole $g$ factor, $\Delta g_h^{*}$, which leads to a dephasing of the
hole spin polarisation due to different precession frequencies. The
dephasing rate due to this inhomogeneity is proportional to the
applied magnetic field. Therefore, the hole spin dephasing time
observed in the experiment, which is the spin dephasing time of an
inhomogeneously broadened ensemble, $T_2^*$, is in first approximation given by~\cite{Bayer_Book}
\begin{equation}
T_2^*=\left (\frac{1}{T_2}+\frac{\Delta g_h^{*} \mu_B
B}{\hbar}\right )^{-1},
\label{T2}
\end{equation}
if  $\Delta g_h^{*}$ is considered as the only source of
inhomogeneity. Here, $T_2$ is the hole spin dephasing time in the
absence of inhomogeneous broadening.

In order to understand the influence of the QW width on the hole
SDT, we have to think about the main mechanisms for hole spin
dephasing. The HH and LH states have different angular momenta,
transitions between these states will therefore destroy hole spin
orientation. In bulk GaAs, where HH and LH valence bands are
degenerate at $k=0$, any momentum scattering may lead to a
transition between HH and LH states, which leads to hole SDTs on the
order of the momentum scattering time \cite{Hilton02}. In QWs, this
degeneracy is lifted. However, for $k~>~0$, the valence bands have a
mixed HH/LH character \cite{Pfalz}, which may also lead to rapid
hole spin dephasing due to scattering \cite{Damen91}. At low
temperatures, resident holes in QWs may become localised at
potential fluctuations within the QW, arising from QW thickness
fluctuations due to monolayer steps at the interfaces, as well as
from the granular distribution of the remote donors. Localisation
significantly reduces the hole quasimomentum, keeping resident holes
in HH states with $k \sim 0$. However, even for $k \sim 0$, there is
a finite admixture of the LH states to the first HH subband
\cite{Luo09}. The significant increase in hole SDT with decreasing
QW width can therefore be attributed to an increased HH/LH
splitting, which reduces the LH contribution to the HH ground state
\cite{WinklerSpringer}. In a first approximation, if we consider
infinitely high square-well potentials, this energy splitting
$\Delta E$ is proportional to 1 over the well-width $L$ squared:
\begin{equation}
\Delta E \sim
\left(\frac{1}{m_{LH}}-\frac{1}{m_{HH}}\right)\frac{\hbar^2
\pi^2}{2L^2}.
\end{equation}
For narrower QWs of finite potential height, however, the hole wave
function penetrates strongly into the barrier, and the energy
splitting is reduced again. For a QW width of 4~nm, the maximum
HH/LH splitting was theoretically predicted and experimentally
observed \cite{Khalifi89}. As figure~\ref{Vgl_Width_2Panel}(b) shows, we
indeed observe the longest hole SDT in sample D with a well width of
4~nm. This suggests that the maximum HH/LH splitting in this sample
is responsible for the long SDT. We will discuss the possible
alternative mechanisms, which might govern hole spin dephasing in
the last section. Before, we will in the following explore the
limits of hole spin dephasing in our samples, starting with the
optical initialisation process of hole spin polarisation in the next
section.

\section{Buildup of a resident hole spin polarisation}
\begin{figure}
\centering
  \includegraphics[width= 0.8\textwidth]{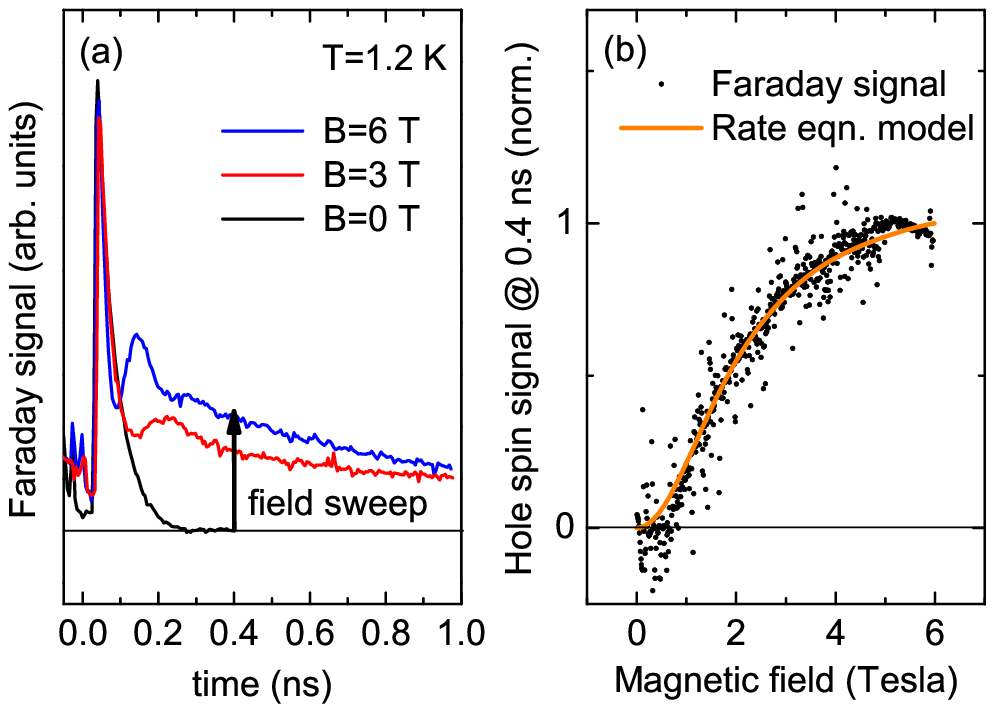}
  \includegraphics[width= 0.6\textwidth]{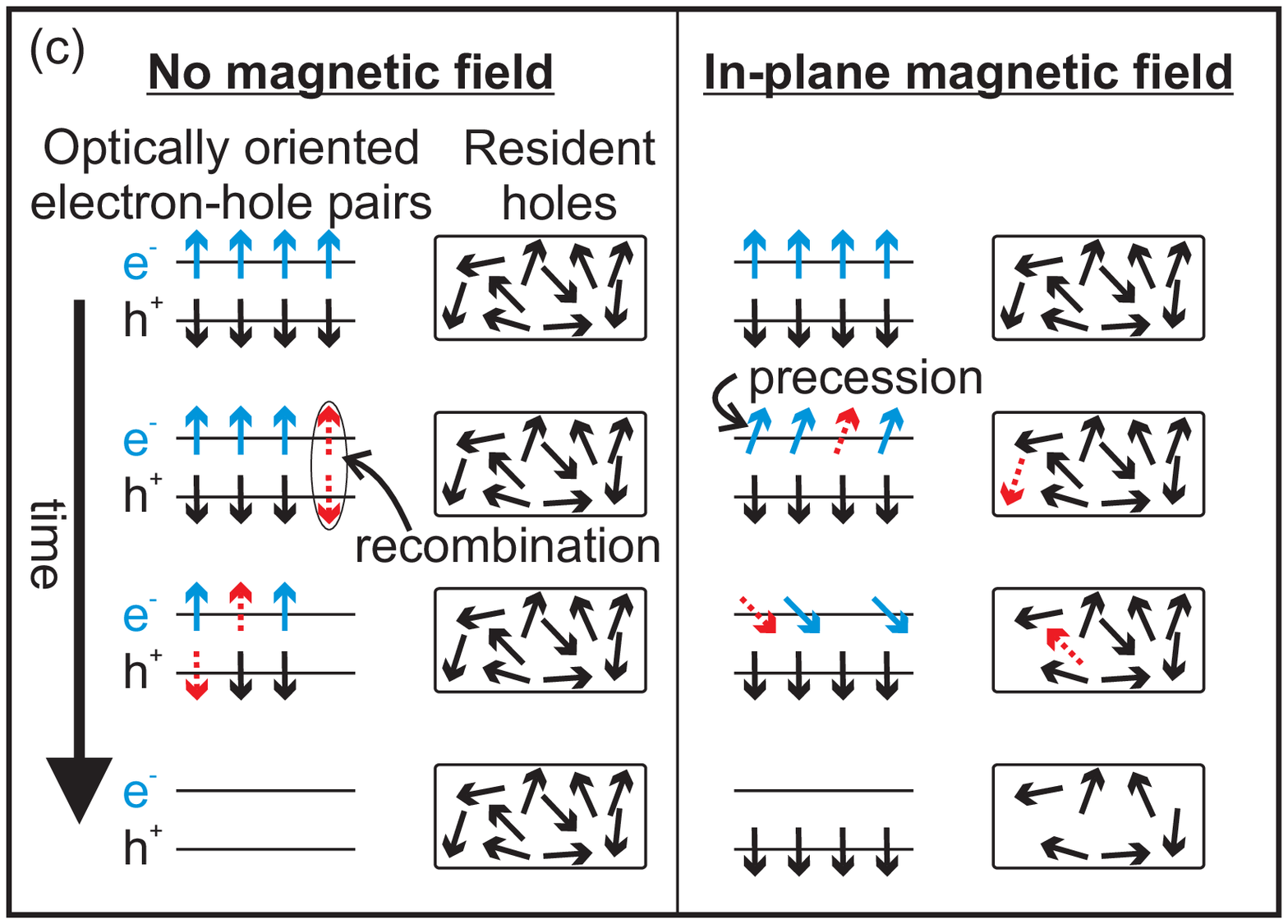}
  \caption{{\bf TRFR for different magnetic fields and buildup of hole spin polarisation.} (a) TRFR traces for sample C taken at 1.2~K for different magnetic fields. The arrow indicates the delay position for the magnetic field sweep shown in (b).
   (b) Kerr signal (black dots) as a function of magnetic field taken at a fixed time delay (400~ps) between pump and probe pulses. The orange line shows the buildup of the hole spin polarisation as a function of magnetic field as calculated by the rate equations model. (c) Diagram of the combined spin and recombination dynamics with (right panel) and without an applied magnetic field (left panel).}
   \label{Buildup}
\end{figure}

In this section, we present experimental results and simulation data concerning the initialization  process of the hole spin polarisation, which was described by Syperek et al.~\cite{syperek}: excitation of the sample with circularly-polarised light will create spin-polarised
electron-hole pairs according to the optical selection rules. If there is no significant electron spin dephasing during the
photocarrier lifetime, the optically oriented electrons will recombine with holes which have matching spin orientation.
Therefore, no spin polarisation will remain within the sample after photocarrier recombination. This process is sketched schematically in the left panel of figure~\ref{Buildup}(c).

In an applied in-plane magnetic field, however, electrons and holes
precess with  different precession frequencies due to their
strongly different $g$ factors. Therefore, during their photocarrier
lifetime, the electron spins will recombine with holes with
arbitrary spin orientation. A part of the optically oriented holes
may therefore remain within the sample after photocarrier
recombination, as depicted in the right panel of figure~\ref{Buildup}(c).

In order to gain deeper insight into the combined dynamics of
electron and hole spins, and to analyse our experimental results
quantitatively, we set up a model in the following. The combined
dynamics of the electron and hole spins can be described via coupled
differential equations for the electron (\textbf{e}) and hole
(\textbf{h}) spin polarisation vectors:
\begin{eqnarray}
  \frac{d \textbf{e}}{dt} &=& -\frac{\textbf{e}}{\tau_R} + \frac{g_e \mu_B}{\hbar} (\textbf{B}\times\textbf{e})\\
  \frac{d \textbf{h}}{dt} &=& -\frac{\textbf{h}}{\tau_h}  + \frac{g_h \mu_B}{\hbar} (\textbf{B}\times\textbf{h}) + \frac{e_z \textbf{z}}{\tau_R}
\end{eqnarray}
In our model we assume the following: the electron spin polarisation
is reduced at a rate, given by the photocarrier recombination and
precesses about the in-plane magnetic field vector \textbf{B}.
Electron spin dephasing has been neglected, as the electron SDT is
expected to be significantly longer than the photocarrier
recombination time $\tau_R$, which is a reasonable assumption. The
hole spin polarisation is reduced at a rate given by the hole SDT
$\tau_h$ and precesses about the in-plane magnetic field vector
\textbf{B}. The last term in the second equation describes the
change of hole spin polarisation due to recombination of
spin-polarised electrons with holes with matching spin orientation.
Here, $e_z$ is the $z$ component of the electron spin polarisation,
and $\textbf{z}$ is the unit vector along the growth direction. A similar approach was used by Yugova et al.~\cite{Yakovlev_Trion} to describe  the initialization of a
resident \emph{electron} spin polarisation in n-doped QWs. However, we need to include, both, electron and hole spin precession (second term in equation (5)) in our differential equations to correctly model the resonant spin amplification spectra we observe, as described in the section below.
In
order to test the validity of this model, we compare it to
experimental results. Figure~\ref{Buildup}(a) shows TRFR traces of
sample C, taken at different applied in-plane fields. The sample has
\begin{figure}
\centering
  \includegraphics[width= 0.8\textwidth]{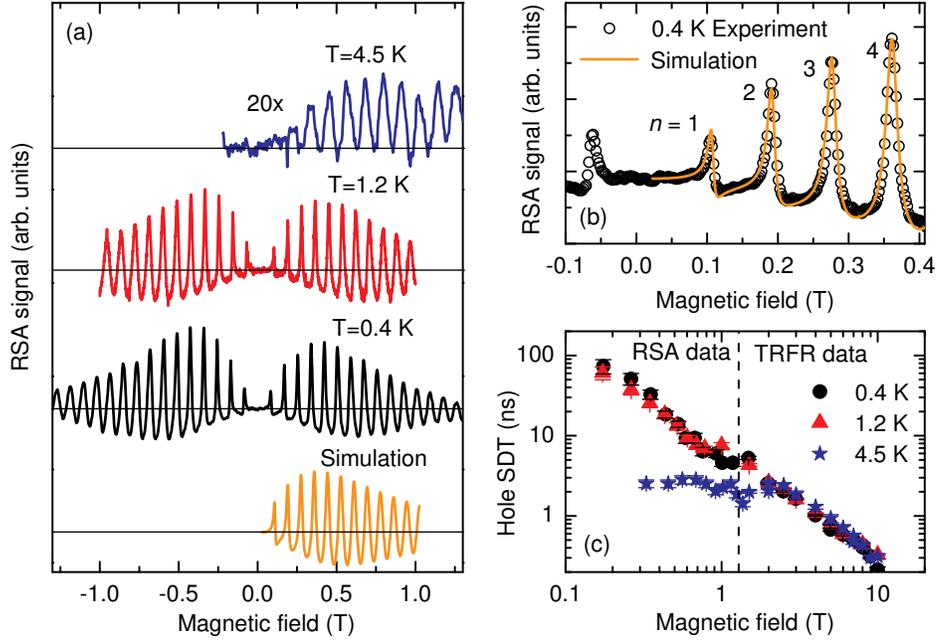}
   \caption{{\bf RSA measurements for different temperatures and magnetic field dependence of hole SDT for different temperatures.}
   (a) RSA traces for sample D, measured at different temperatures, compared to simulation data.
   (b) RSA trace measured at 400~mK (open circles) compared to simulation data (orange line). (c) Hole SDT,
   determined from RSA data (left of the dotted line and TRFR data (right of the dotted line), for different
   temperatures as a function of magnetic field.}
   \label{RSA_3Panel}
\end{figure}been carefully aligned so that the magnetic field is applied along
the QW plane, i.e., $\alpha \sim 0$, resulting in a very low
effective hole $g$ factor $g_h^{*}$. For zero magnetic field, the
TRFR signal decays monoexponentially with a decay constant of
$\tau_R$= 48~ps, which reflects the photocarrier recombination time in sample C.
For larger magnetic fields, pronounced electron spin precession can
be observed during $\tau_R$, and a significant nonzero TRFR signal
remains after photocarrier recombination. This signal is due to the
buildup of a resident hole spin polarisation, as explained above.
Its amplitude increases as the magnetic field is increased. Due to
the very small hole $g$ factor ($|g_h^{*}|<0.005$), no hole spin
precession is observed in the time range shown in
figure~\ref{Buildup}(a). This allows us to study the buildup of the hole
spin polarisation in more detail using the following experimental
technique: the TRFR signal is recorded for a fixed time delay
between pump and probe. The time delay is chosen such that
photocarrier recombination is complete, thereby leaving a TRFR
signal from resident holes, only. The magnetic field is ramped up
from zero in small increments. Figure~\ref{Buildup}(b) shows such a
measurement for a time delay of $\Delta t=400$~ps. It is clearly
visible how the resident hole spin polarisation increases from zero
to a maximum value, at which it saturates. This buildup can easily
be modeled by the differential equation system described above. For
this, we used the parameters extracted from the experiment:
$|g_e|=0.118$, $\tau_R$= 48~ps, $g_h^{*}=0$. Hole spin dephasing was
neglected to model the dataset. The resulting hole spin polarisation
as a function of magnetic field is shown as an orange solid line in
figure~\ref{Buildup}(b). It is in excellent agreement with the
experimental data. These measurements clearly demonstrate the
crucial role of an applied in-plane magnetic field for transferring
spin polarisation from optically oriented photocarriers to resident
holes.

\section{Resonant spin amplification measurements of hole
spin dynamics} To explore the hole SDT in very small applied
magnetic fields, and to approach the ultimate limit of hole SDT in
our samples, we employ the RSA technique \cite{Kikkawa98}, which has
been used in recent years, e.g., to study electron spin dynamics in
n-doped bulk GaAs \cite{Kikkawa99} and 2DES in CdTe-based QWs
\cite{Yakovlev_RSA}. RSA is based on the constructive interference
of spin polarisations created by subsequent laser pulses in a
time-resolved Faraday or Kerr rotation measurement. The Faraday/Kerr
signal is measured as a function of the in-plane magnetic field for
a fixed time delay $\Delta t$, typically before the arrival of a
pump pulse. If the optically oriented spins precess by an integer
multiple of $2\pi$ within the time delay between two pump pulses, a
maximum in the RSA signal is observed. The maxima are typically
periodic in $B$. The SDT can be determined from the half-width of
the maxima, and their spacing yields the $g$ factor. In systems
where the buildup of a resident spin polarisation is governed by the
interplay of electron and hole spin dynamics, however, a more
complex shape of the RSA signal can be expected, as will be
demonstrated now.

Figure~\ref{RSA_3Panel}(a) shows a series of RSA traces, measured on
sample D for different temperatures, compared to a numeric
simulation, using the coupled differential equation model introduced
above. The unusual 'butterfly' shape \cite{Yakovlev_Trion} of the
RSA signals is clearly visible for the traces measured at 1.2~K and
0.4~K. It is well-reproduced by the simulation data. The RSA signal
at 4.5~K is about 20 times weaker than the lower-temperature
signals. In figure~\ref{RSA_3Panel}(b), a closeup of the first few RSA
maxima measured at 0.4~K is compared to simulation data. There is no
RSA maximum in the measurement and the simulation for $B=0$, in
contrast to typical RSA curves measured, e.g., in n-doped GaAs bulk.
This is due to the peculiar transfer process that leads to a
resident hole spin polarisation. As demonstrated above, a finite
in-plane magnetic field is necessary to create a resident hole spin
polarisation. The first maximum at finite field (numbered as $n=1$)
has a distinct shape resembling the derivative of a Lorentzian,
while subsequent maxima resemble slightly asymmetrical Lorentzian
curves. The amplitude of the maxima first increases with $n$, then
decreases again for $n >5$ (see figure~\ref{RSA_3Panel}(a)), while the
FWHM of the maxima increases. These features are clearly reproduced
in the simulation. They can be explained as follows: for small
magnetic fields, there is only a partial transfer of spin
polarisation to the resident holes, a process which saturates in our
experiment for $B\sim 0.5$~T. For larger magnetic fields, the RSA
FWHM increases due to inhomogeneous broadening, and the amplitudes
of the RSA maxima decrease accordingly, until the RSA signal drops
to the noise level at about 1.6~T in the low-temperature
measurements. Interestingly, direct comparison of the RSA traces measured at 1.2~K and 4.5~K reveals that the effective hole g factor $|g_h^{*}|$ decreases from 0.067 to 0.050 as the temperature is increased, in good agreement with previous observations~\cite{syperek}.

Figure~\ref{RSA_3Panel}(c) shows the hole spin lifetime in sample D
for different temperatures. The data for low magnetic fields (below
1~T) have been determined from RSA data, the data for higher
magnetic fields (above 2~T) were determined from TRFR data.
Interestingly, while at 4.5~K, the hole SDT saturates at about
2.5~ns, even in low magnetic fields, it follows a clear $1/B$-like
dependence down to very small magnetic fields in the data measured
at lower temperatures, yielding values of $T_2^*=74 \pm 15$~ns at
0.4~K and $T_2^*=61 \pm 11$~ns at 1.2~K (determined from the FWHM of
the second RSA maximum at $B\sim 0.2$~T) \cite{foot2}. The power of
the RSA technique is evident if one compares the hole SDT measured
at higher fields: in the magnetic field range where TRFR
measurements yield precise results of the hole SDT, very little
difference is observed in the values of the hole SDT in the
temperature range from 0.4~K to 4.5~K. From the magnetic field
dependence of the hole SDT, we may infer that the $T_2$ time of the
hole spin is above 80~ns at temperatures below 500~mK. By fitting equation~\ref{T2} to the magnetic field dependence, we determine the g factor inhomogeneity $\Delta g_h^{*} = 0.003 \pm 0.0002$.

\begin{figure}
\centering
  \includegraphics[width= 0.8\textwidth]{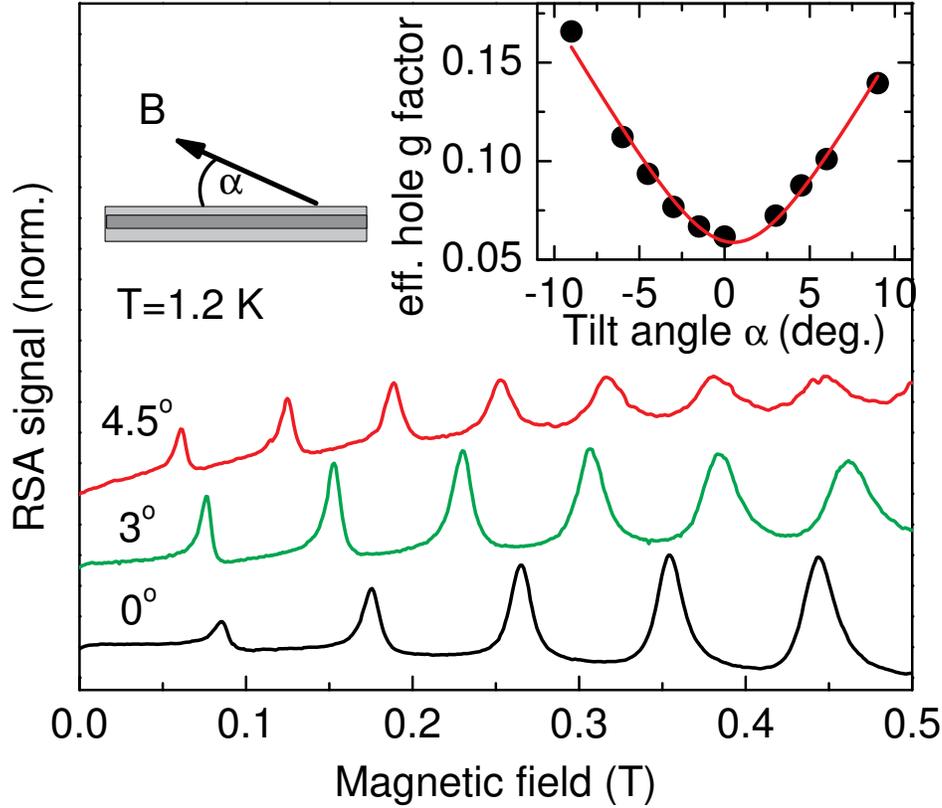}
   \caption{{\bf RSA measurements for different tilt angles of the magnetic field and effective hole g factor dependence on tilt angle.} RSA traces for sample D, measured at 1.2~K, for different tilt angles of the sample with respect to the external magnetic field.
   The inset shows the effective hole g factor $|g_h^{*}|$ determined from the spacing of the RSA maxima (black dots). The solid red line represents a fit of the results using equation \ref{GeoSum}.}
   \label{RSA_Winkel}
\end{figure}

Furthermore, the RSA measurements allow for a precise determination
of the effective hole $g$ factor, which is given by $g_h^{*}=2 \pi
f_{rep} \hbar/(\mu_B \Delta B)$. Here, $f_{rep}$ is the laser pulse
repetition frequency, and $\Delta B$ is the magnetic field spacing
between two adjacent RSA maxima. Figure~\ref{RSA_Winkel} shows RSA
traces measured at 1.2~K for different angles $\alpha$ of the
magnetic field with respect to the QW plane~\cite{foot3}. The geometry is
sketched in the figure. The spacing of the maxima is significantly
reduced as the magnetic field acquires an out-of-plane component.
The effective hole $g$ factor $|g_h^{*}|$ is extracted from this spacing, the
results are shown in the inset. The increase of the effective hole
$g$ factor with the magnetic field angle is due to an admixture of
the out-of-plane component of the hole $g$ factor ${g}_{||}$, which is
typically far larger than the in-plane hole $g$ factor ${g}_{\bot}$, as described by equation~\ref{GeoSum}. By fitting the results
with this equation as indicated by the solid red line in the inset, both components of the hole g factor at 1.2~K can be determined with high accuracy: $|{g}_{\bot}|= 0.059 \pm	0.003$,
$|{g}_{||}|=0.89 \pm 0.03$.

Finally, we will discuss the possible mechanisms, which might limit
the hole SDT in our experiments. At low temperatures and in low
magnetic fields, where the $g$ factor inhomogeneity may be
neglected, in principle several mechanisms might be thought of to
pose the ultimate limit of the hole SDT. In the following, we will
discuss their dependence on QW thickness in the light of our
experiments, and identify the most important mechanism.

\paragraph{Thermal activation of holes from localized states:}
The thermal activation rate is proportional to the Boltzmann factor
  $exp\left((-E_{Loc)}/(k_B T)\right)$. For localization of holes
  at QW monolayer thickness fluctuations, the localization energy
  increases drastically with a reduction of the QW width \cite{Luo09}.
  For a 15~nm wide QW, it is less than 1~meV, while for a 4~nm wide QW,
  it is about 10~meV. In both cases, the localization energy is
  significantly larger than the thermal energy, even at the highest
  measurement temperature used in our experiments: $E_{th}$(4.2~K)=0.36~meV.
  This indicates that thermal activation of holes out of localized states
  is unlikely to affect hole spin dynamics at liquid-helium temperatures and below.

\paragraph{Dipole-Dipole interaction of holes with nuclei:}
Even though the p-like symmetry of valence band states does not
allow for a \emph{contact}
  hyperfine interaction between holes and nuclei, a coupling is possible via
  dipolar interactions \cite{Loss08,Marie09}. The nuclei act as slowly
  varying random magnetic fields, which lead to both, single spin decoherence
  and ensemble spin dephasing. The variance of the nuclear magnetic field is
  inversely proportional to the number of nuclei that interact with the hole
  confined in the quantum dot, which may be estimated from the dot
size. We note that the typical size of quantum dots which arise from
monolayer thickness fluctuations in high-quality GaAs QWs (TF-QDs)
is on the order of (100~nm)$^2$, significantly larger than that of
self-assembled InAs dots. One may estimate that a hole interacts
with about 10$^6$ nuclei in TF-QDs or electrostatically controlled
QDs \cite{Petta05}, whereas  self-assembled dots typically contain
only about $5 \cdot 10^4$ nuclei \cite{Marie09}. Additionally, the
random magnetic fields associated with the nuclei may be suppressed
very efficiently by a small, external  magnetic field applied within
the sample plane \cite{Loss08}. It is shown above that the
application of an in-plane magnetic field is necessary for the
transfer of spin polarisation of optically oriented carriers to
resident holes. Therefore, we may neglect hole spin decoherence due
to interaction with nuclei as the dominant process, limiting the
hole SDT in our experiment.

\paragraph{Finite admixture of LH states to HH states:}
Even localized holes have a finite quasimomentum, which is  given by
the hole temperature. In an analogon to the Elliott-Yafet
\cite{Elliot54,Yafet63} mechanism, which has been studied in detail
for electrons, any momentum scattering  may therefore destroy hole
spin orientation via hole spin flip due to the slight admixture of
LH states to HH states. This admixture is on the order of 2~percent
for thin QWs \cite{Luo09}. It increases significantly for wider QWs
due to the decreasing HH/LH energy splitting, increasing the
probability of a hole spin flip during momentum scattering
\cite{Wu05}. The spin dephasing rate is directly proportional to the
momentum scattering rate, which in turn increases with the sample
temperature.

Summarizing this part: We believe that the increased HH/LH splitting
is responsible for the increased hole SDT with decreasing QW width
in our experiments. On the other hand, the still finite LH admixture
to the HH ground state and, hence, the possibility of spinflip
scattering of Elliott-Yafet type sets the limit for hole spin
dephasing.

\section*{Conclusions}
In conclusion, we have investigated hole spin dynamics in
two-dimensional hole systems embedded in quantum wells of different
width. Due to the increased energy splitting between heavy- and
light-hole states, the hole spin dephasing time increases
drastically in narrow quantum wells. The inhomogeneity $\Delta
g_h^{*}$ of the hole $g$ factor leads to a typical $1/B$-like
dependence of the hole SDT. In order to transfer spin polarisation
from the optically oriented photocarriers to the 2DHS, a finite
magnetic field is necessary. This also leads to a peculiar shape of
resonant spin amplification signals measured on 2DHS. From these RSA
signals, the hole SDT in low magnetic fields can be extracted. It is
on the order of 80~ns, one order of magnitude larger than for
electrons in quantum dots of similar dimensions defined in 2DES by external gates~\cite{Petta05}. This makes 2DHS an interesting material
system for scalable quantum computing devices based on electrostatically confined charge carriers. Additionally, RSA measurements in tilted magnetic fields allow us to accurately determine both, the in-plane and out-of plane components of the hole g factor.

\ack
The authors would like to thank E.L. Ivchenko, M.M. Glazov and M.W. Wu for fruitful discussion.  Financial support by the DFG via SPP 1285 and SFB 689 is gratefully
acknowledged.


\begin{thebibliography}{0}
\bibitem{foot3}
The sample tilt angle is determined by measuring the angle of the pump beam which is reflected from the sample surface.
\bibitem{foot2}
The hole SDT is proportional to $1/(\delta B^2)$ for all RSA maxima with $n>1$. Here, $\delta B$ is the full width at half maximum (FWHM) of the RSA maximum.
\end{thebibliography}
\end{document}